\documentclass{emulateapj}
\usepackage{amssymb}   
\usepackage{colordvi}   

\shorttitle{The IMF at at 0.9$<z<$1.5} \shortauthors{I.
  Mart\'in-Navarro et al. }

\def\gsim{ \lower .75ex \hbox{$\sim$} \llap{\raise .27ex \hbox{$>$}} }
\def\lsim{ \lower .75ex \hbox{$\sim$} \llap{\raise .27ex \hbox{$<$}} }

\begin{document}
\title{The stellar initial mass function at 0.9$<$\lowercase{z}$<$1.5}

\author{Ignacio Mart\'in-Navarro$^{1,2}$, Pablo G. P\'erez-Gonz\'alez$^{3,4}$, 
Ignacio Trujillo$^{1,2}$, Pilar Esquej$^{3}$, Alexandre Vazdekis$^{1,2}$, Helena 
Dom\'{\i}nguez S\'anchez$^{3}$, Guillermo Barro$^{5}$, Gustavo Bruzual$^{6}$, 
St\'ephane Charlot$^{7}$, Antonio Cava$^{8}$, Ignacio Ferreras$^{9}$, N\'estor 
Espino$^{3}$, Francesco La Barbera$^{10}$, Anton M. Koekemoer$^{11}$, A. Javier Cenarro$^{12}$}
\affil{$^{1}$Instituto de Astrof\'{\i}sica de Canarias,c/ V\'{\i}a L\'actea s/n, E38205 - La Laguna, Tenerife, Spain}
\affil{$^{2}$Departamento de Astrof\'isica, Universidad de La Laguna, E-38205 La Laguna, Tenerife, Spain} 
\affil{$^{3}$Departamento de Astrof\'{\i}sica, Facultad de CC. F\'{\i}sicas, Universidad Complutense de Madrid, E-28040 Madrid,  Spain} 
\affil{$^{4}$Severo Ochoa Visitor at Instituto de Astrof\'{\i}sica de Canarias} 
\affil{$^{5}$UCO/Lick Observatory, Department of Astronomy and Astrophysics, University of California, Santa Cruz, CA 95064, USA} 
\affil{$^{6}$Centro de Radioastronom\'{\i}a y Astrof\'{\i}sica, UNAM, Campus Morelia, M\'exico} 
\affil{$^{7}$UPMC-CNRS, UMR7095, Institut d'Astrophysique de Paris, F-75014 Paris, France} 
\affil{$^{8}$Observatoire de Gen\`eve, Universit\'e de Gen\`eve, 51 Ch. des Maillettes, 1290, Versoix, Switzerland} 
\affil{$^{9}$Mullard Space Science Laboratory, University College 
London, Holmbury St Mary, Dorking, Surrey RH5 6NT} 
\affil{$^{10}$INAF - Osservatorio Astronomico di Capodimonte, Napoli, Italy}
\affil{$^{11}$ Space Telescope Science Institute, 3700 San Martin Drive, Baltimore, MD 21218, USA}
\affil{$^{12}$ Centro de Estudios de F\'{\i}sica del Cosmos de Arag\'{\o}n, Plaza San Juan 1, 44001 Teruel, Spain}
\email{email: imartin@iac.}

\begin{abstract}

  We explore the stellar initial mass function (IMF) of a sample of 49
  massive quiescent galaxies (MQGs) at 0.9$<$z$<$1.5. We base our
  analysis on intermediate resolution spectro-photometric data in the
  GOODS-N field taken in the near-infrared and optical with the
  HST/WFC3 G141 grism and the Survey for High-z Absorption Red and
  Dead Sources (SHARDS). To constrain the slope of the IMF, we have
  measured the TiO$_2$ spectral feature, whose strength depends
  strongly on the content of low-mass stars, as well as on stellar
  age.  Using ultraviolet to near-infrared individual and stacked
  spectral energy distributions, we have independently estimated the
  stellar ages of our galaxies. Knowing the age of the stellar
    population, we interpret the strong differences in the TiO$_2$
    feature as an IMF variation. In particular, for the heaviest
  z$\sim$1 MQGs (M$>$10$^{11}$M$_\sun$) we find an average age of
  1.7$\pm$0.3~Gyr and a bottom-heavy IMF ($\Gamma_b$=3.2$\pm$0.2).
  Lighter MQGs ( 2$\times$10$^{10}$$<$M$<$10$^{11}$~M$_\sun$) at the same
  redshift are younger on average (1.0$\pm$0.2~Gyr) and present a
  shallower IMF slope ($\Gamma_b=2.7^{+0.3}_{-0.4}$). Our results are
  in good agreement with the findings about the IMF slope in
  early-type galaxies of similar mass in the present-day Universe.
  This suggests that the IMF, a key characteristic of the stellar
  populations in galaxies, is bottom-heavier for more massive galaxies
  and has remained unchanged in the last $\sim$8 Gyr.

\end{abstract}


\section{Introduction}

The initial mass function (IMF) dictates the distribution of stellar
masses for any single star formation event in a galaxy.  Consequently,
it determines the number of massive stars formed and being responsible
for the feedback and chemical processes. The IMF also fix the numbers
of low-mass stars, which dominate the total stellar mass of a galaxy.

Growing evidence support a non-universal IMF in the nearby Universe,
where massive early-type galaxies (ETGs) show an enhanced fraction of
dwarf stars in the center compared to the Milky Way \citep{vandokkum}.
Moreover, the dwarf-to-giant ratio, i.e., the IMF slope, correlates
with the central velocity dispersion
\citep{cenarro,treu,cappellari,ferreras,labarbera,Conroy13,Spiniello2013}.
These results challenge the existence of a universal IMF inferred from
resolved stellar population analysis in the Local Group
\citep{kroupa,bastian,kroupa13}.

To have a consistent picture of galaxy evolution, it is absolutely
necessary to investigate the IMF at different redshifts. So far, the
IMF of z$\lesssim$1 galaxies has been studied indirectly using virial
masses \citep[][]{renzini,vds13} or elaborated dynamical models
\citep{Shetty}. These works point to a \citet{salpeter} IMF for
massive galaxies at intermediate redshift.  Other indirect
IMF-sensitive observables have also been used in the topic. For
instance, the consistency between the cosmic stellar mass and star
formation rate densities \citep{dave,pablo08} and the luminosity
evolution of massive ETGs \citep{vd08} are better described by a
flatter (i.e., with a relatively larger number of massive stars) IMF
at higher look-back times. Even in star-forming galaxies, the
  constancy of the IMF is in tension with observations
  \citep{Hoversten,Meurer09}

Here we explore, for the first time, the IMF slope at z$\gtrsim$1
using stellar populations synthesis models in massive quiescent
galaxies (MQGs). To achieve this goal, we study the TiO$_2$
IMF-sensitive spectral feature \citep{tio2}. In Section 2, we describe
the data. The IMF inference is explained in Section~3. In Section~4,
we discuss our results. We adopt a standard cosmology: H$_0$= 70 km
s$^{-1}$ Mpc$^{-1}$, $\Omega_m$= 0.3, and $\Omega_\Lambda$=0.7.

\section{Sample and data description}
\label{sec:data}

\begin{figure}
  \begin{center}
    \includegraphics[scale=0.95,bb= 150 194 450 633,clip,width=0.5\textwidth]{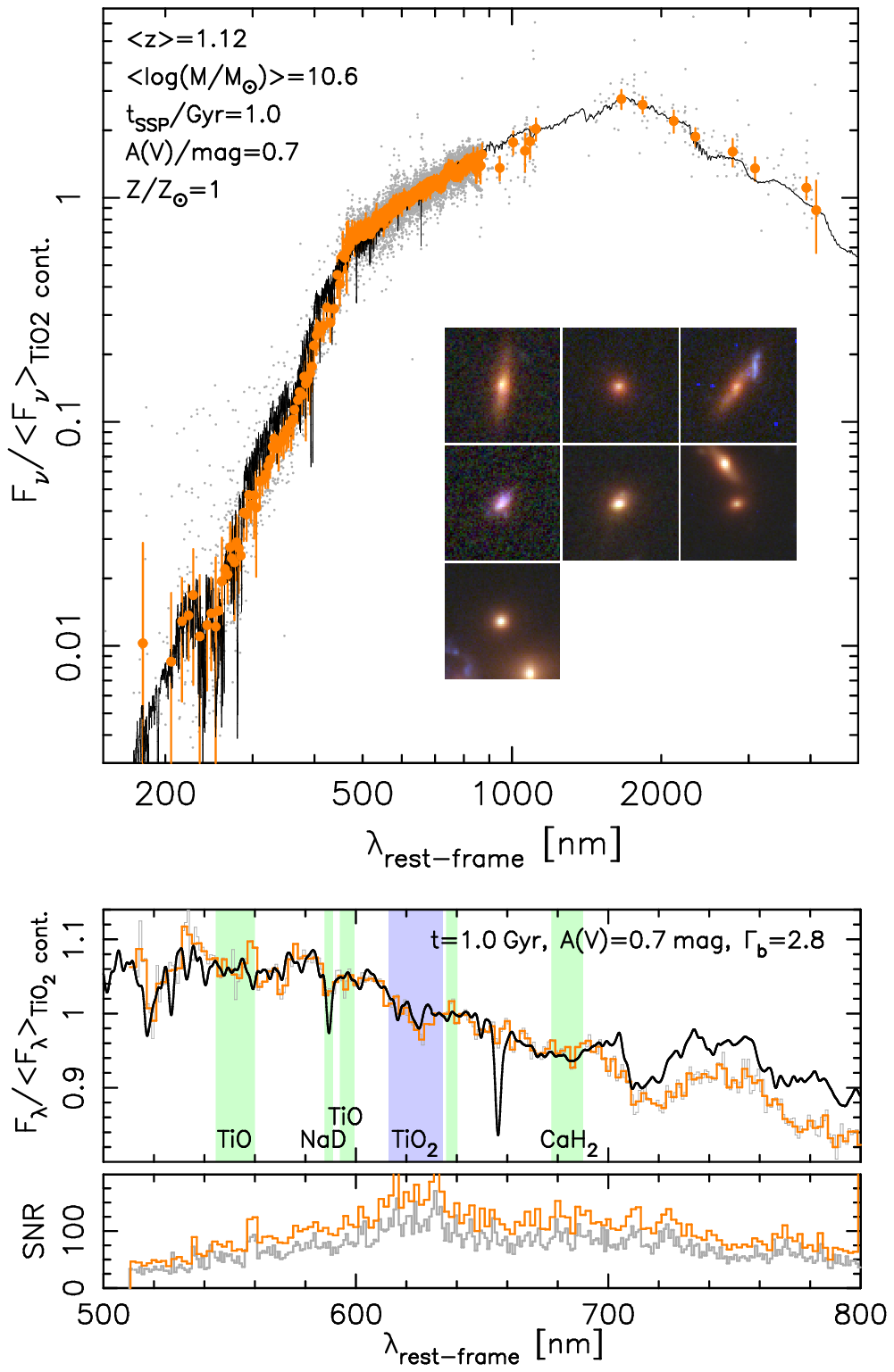} 
    \caption{Stacked SEDs (normalized to the average TiO$_2$ continuum
      flux) for MQGs at 0.9$<$z$<$1.5 in GOODS-N for the low-mass
      sample. We show the complete UV-to-NIR stack on {\it top}, with
      data for individual galaxies (gray dots) and average fluxes in
      bins of 20 photometric data points (orange), including 2$\sigma$
      bars.  The black line shows best-fitting SSP models
      (BC03/XMILES, Kroupa IMF, \citet{Calzetti} attenuation law).
      We provide 5''$\times$5'' RGB postage stamps for representative examples of the sample. At
      the {\it bottom}, we show the WFC3/G141 grism data including
      stacked (gray) and smoothed (orange) spectra (using 10 and
      20~\AA\, bins, respectively), and their SNR. The black line
      shows best-fitting MIUSCAT SSP models. Shaded regions mark the
      TiO$_2$ absorption (blue), and other IMF-sensitive indices (green). 
      Deviations from an SSP appear beyond 700nm, where 
      a small fraction ($\sim10$\%) of a younger population can significantly
      affect the continuum, but barely changes (0.004 mag) the TiO$_2$ value.}
  \label{data}
  \end{center}
\end{figure}

\begin{figure}
  \begin{center}
    \includegraphics[scale=0.95,bb= 150 194 450 633,clip,width=0.5\textwidth]{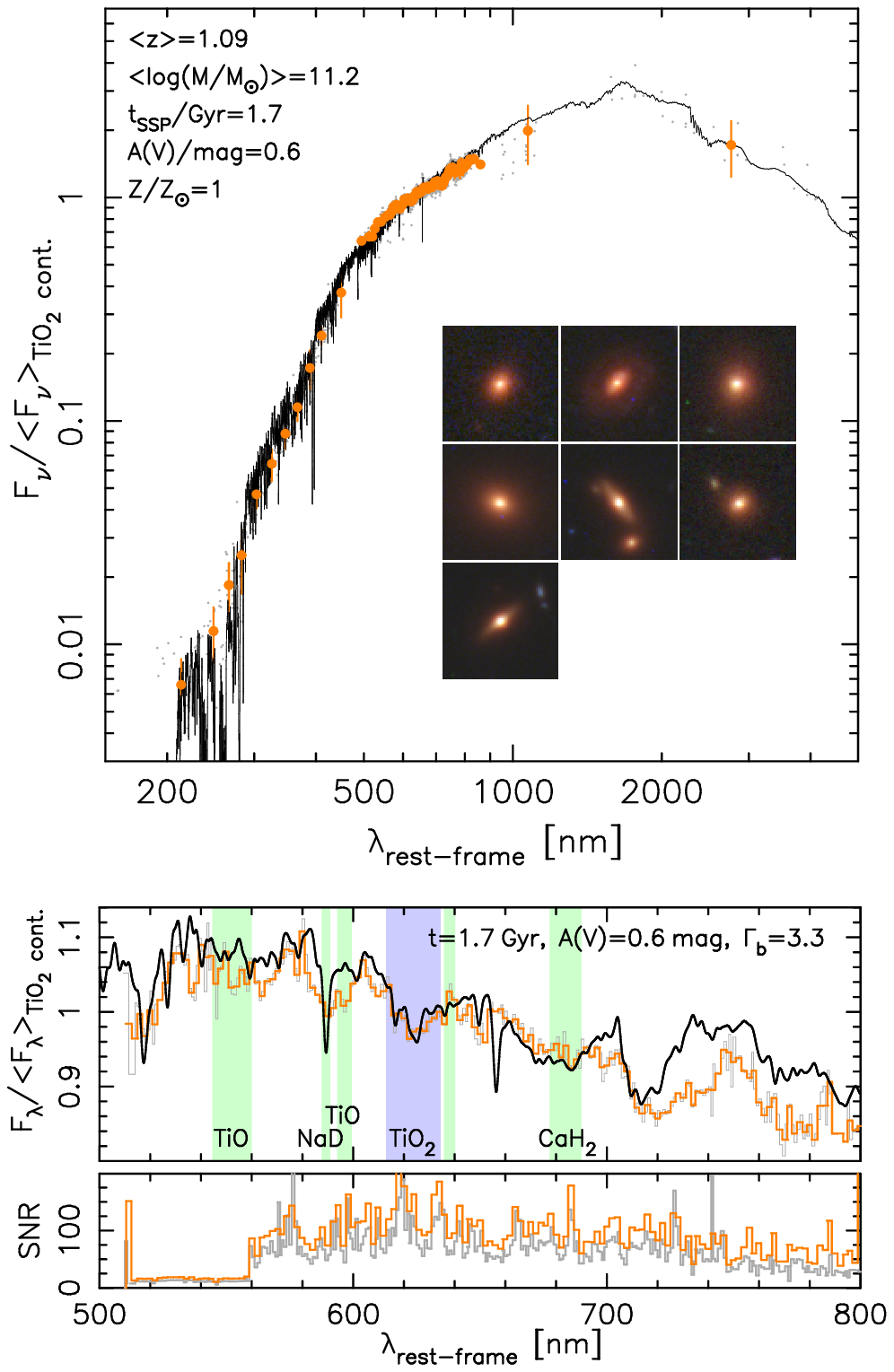} 
    \caption{Same as Fig.~\ref{data} but for the high-mass sample.}
  \label{data2}
  \end{center}
\end{figure}


To facilitate the determination of the IMF slope at high-z, we study
galaxies with no signs of recent star formation (quiescent galaxies).
These objects have simpler Star Formation Histories (SFHs) than
star-forming galaxies and are sufficiently well represented by a
single stellar population (SSP) model \citep[e.g.,][]{Whitaker}.

MQGs at 0.9$<$z$<$1.5 were selected with two criteria: (1) the $UVJ$
diagram complemented with fluxes in the MIR/FIR; and (2) a sSFR vs.
stellar mass plot. We worked with the mass selected sample presented
in \citet{pablo08}.  From this work, we took the spectral energy
distributions (SEDs), stellar population and dust emission models for
all IRAC sources in GOODS-N.  Those SEDs were complemented with
medium-band optical photometry from the Survey for High-z Absorption
Red and Dead Sources, SHARDS \citep{shards}. The broad- and
medium-band photometry was fitted with a variety of stellar population
models to obtain photometric redshifts, stellar masses, SFRs, and
rest-frame synthetic colors \citep[see][]{barro11a,barro11b}.  Thanks
to the ultra-deep medium-band data from SHARDS, the quality of our
photometric redshifts is excellent: the median $\Delta z/(1+z)$ is
0.0067 for the 2650 sources with $I$$<$25 (P\'erez-Gonz\'alez \emph{et
  al.} 2014, in prep; \citealt{ferreras14}). SFRs were calculated for
all galaxies using various dust emission templates and the {\it
  Spitzer}-MIPS and {\it Herschel}-PACS/SPIRE fluxes, jointly with
UV-based measurements for non-detections in the MIR/FIR. The UV-based
SFRs were corrected for extinction with the UV slope $\beta$ and an
extrapolation of the IR-$\beta$ (IRX) relationship \citep{Meurer}. The
extrapolation technique was developed to recalibrate the IRX-$\beta$
relation using faint IR emitters (more similar to MIR-undetected
galaxies) at the same redshifts. Details about the selection will be
given in Dom\'{\i}nguez S\'anchez \emph{et al.}  (2014, in prep.).

Using this dataset, we selected galaxies at 0.9$<$z$<$1.5 having
stellar masses M$>$ 2$\times$10$^{10}$~M$_\sun$ (\citealt{mw} IMF), and
rest-frame $UVJ$ colors within the quiescent galaxy wedge
($U$$-$$V$$>$1.3, $V$$-$$J$$<$1.6,
$U$$-$$V$$>$0.88$\times$$(V$$-$$J)$+0.59; \citealt{Whitaker2011}).
The mass cut was chosen to allow measuring the TiO$_2$ absorption in
the grism spectra described below. The $UVJ$-selected sample was
complemented with galaxies with sSFR$<$0.2~Gyr$^{-1}$, our limit for
quiescence. Galaxies with MIPS detections were removed from
  the sample, as the MIR emission indicates active/residual star
formation or nuclear activity, which would complicate the stellar
population analysis.  Using these two criteria, we selected 124
sources in the 112 arcmin$^2$ covered simultaneously by GOODS, SHARDS,
CANDELS, and {\it Herschel}-GOODS.

The TiO$_2$ spectral index was measured in stacked WFC3/G141 grism
data (covering 1.1$\lesssim$$\lambda$$\lesssim$1.6~$\mu$m) from the
AGHAST survey (PI: Weiner).  We selected all galaxies with
$H$$<$25.5~mag from the F160W imaging in CANDELS
\citep{Grogin,Koekemoer} and reduced the grism data to extract 2D
spectra using the aXe software (version 2.3).  Then we collapsed the
data to obtain 1D spectra using our own dedicated software.  The
reduction used 0.064 arcsec/pixel and 23.5~\AA/pixel.  The 1D
extractions were optimized for each galaxy using its effective radius,
position angle, and the contamination map provided by aXe. Visual
inspection helped to remove spectra with significant contamination
and/or artifacts, leaving 97 galaxies with usable G141 spectra. We
kept the spectra with SNR$>$5 per pixel.  Our final sample is composed
by 57 galaxies with  2$\times$10$^{10}$$<$M/M$_\sun$$<$10$^{11.5}$
($<$M$>=$10$^{10.6}$~M$_\sun$) and 0.9$<$z$<$1.5 ($<$z$>=$1.1).
Reliable spectroscopic redshifts were available for 33 galaxies; the
median quality of the photo-redshifts for M$>$10$^{10.5}$~M$_\sun$
galaxies is $\Delta z/(1+z)=$0.0047.

Measurements were carried out in stacked spectra of these 57 z$\sim$1
MQGs. We dissected the sample to probe the lowest and highest mass
regimes with two stacked spectra of similar SNR
(Figures~\ref{data},\ref{data2}). The high-mass sample was composed by
7 galaxies with M$>$10$^{11}$~M$_\sun$ ($H$=19.7-21.3~mag),
and the low-mass spectra by 50 galaxies with M$<$ 2$\times$10$^{10}$~M$_\sun$
($H$=20.3-22.4~mag).  To build the stacks, we first
de-redshifted all individual observed spectra, then normalizing them
to the TiO$_2$ continuum (see next section). We calculated flux
averages and errors in rest-frame wavelength bins of 10~\AA.  Finally,
we smoothed the stacks with a 20~\AA boxcar kernel. The average SNR
per resolution element of the final stacked (smoothed) spectra is 70
(100), 100 (140) around the TiO$_2$ absorption.

\section{SED analysis: ages and IMF slope}

The integrated spectral properties of a SSP are defined by four
parameters: age, metallicity ([Z/H]), IMF and $\alpha$-elements
over-abundance ([$\alpha$/Fe]). In this Letter, we analyze the TiO$_2$
absorption, an IMF-sensitive feature which depends very weakly on
[Z/H] and [$\alpha$/Fe] \citep{thomas11b,labarbera}.  We present the
age and IMF constraints for z$\sim$1 MQGs based on this TiO$_2$
spectral index as well as on the ultraviolet to near-infrared SEDs.
In Section~\ref{sec:dis}, we discuss the impact of the unknown values
of [Z/H] and [$\alpha$/Fe] on our results. The TiO$_2$ absorption is
wide and deep enough to be measured with WFC3 grism data. Measurements
for other IMF-sensitive features (see Figures~\ref{data},\ref{data2})
would be compromised by low SNR at $\lambda_\mathrm{rf}$$<$500~nm, the
low spectral resolution in the case of NaD, or the proximity to
emission features in the case of CaH$_2$. Thus, we concentrate our IMF
analysis on TiO$_2$ measurements.

\subsection{Age determination} \label{sec:age}

To constrain the age of the stellar population, we used three
different methods.  First, we fitted the G141 grism stacked spectra
constrained to the rest-frame wavelength range
500$<$$\lambda_\mathrm{rf}$$<$800~nm (Figures~\ref{data},\ref{data2}).
We used the \citet[][hereafter BC03]{bc03} models fed with the XMILES
library (Charlot \& Bruzual, private communication).  We assumed a SSP
with solar and super-solar metallicities, and a \citet{Calzetti}
attenuation law, and fitted the data to obtain ages, extinctions, and
metallicities.  We tested how the results were affected by: (1) using
\citet{salpeter}, \citet{mw}, and \citet{chabrier} IMFs; (2) different
attenuation recipes, namely, \citet{Calzetti}, appropriate for
starburst galaxies, and the more general law from \citet{Charlot00};
and (3) different stellar population synthesis libraries and codes,
namely, BC03 using XMILES and STELIB\citep{stelib} libraries, and
MIUSCAT \citep{miles}. In all cases, we found negligible differences
in the estimated ages ($<$0.1~Gyr) and extinctions (0.1~mag). Our
fitting method included a Montecarlo algorithm to analyze
uncertainties and degeneracies (see \citealt{shards}). Given the short
wavelength range probed by the grism data, the dust extinction was not
well constrained.  Indeed, we found a strong age-extinction
degeneracy. For example, for the high-mass stack, equally good fits
were obtained for stellar populations with relatively young ages
($\sim$1 Gyr) and large extinctions (A(V)$>$1.5~mag) and for older
ages and lower extinctions (1--2~Gyr and A(V)$<$1~mag). Constraining
the extinction to A(V)$<$1~mag, we found that the stacked high-mass
spectrum was best fitted by a SSP with solar metallicity,
t$=$1.6$\pm$0.2~Gyr, and A(V)$=$0.5$\pm$0.3~mag. The low-mass stack
was best fitted with solar metallicity, t$=$1.0$\pm$0.2~Gyr, and
A(V)$=$0.7$\pm$0.3~mag.

Our second age determination method used the whole UV-to-NIR stacked
SED (Figures~\ref{data}, \ref{data2}). The SHARDS medium-band and
grism data allow accurate measurements of both the 4000~\AA\, break
and the Mg$_\mathrm{UV}$ absorption, two very good age estimators
\citep[see][and references therein]{shards,hernan13,ferreras14}.  The
wider spectral range resulted in better constraints on the age and the
extinction. The best-fitting BC03/XMILES SSP model provided
$t$$=$1.77$\pm$0.17~Gyr, A(V)$=$0.60$\pm$0.06, and
$t$$=$1.02$\pm$0.15~Gyr and A(V)$=$0.70$\pm$0.06 for the high-mass and
low-mass samples, respectively (solar metallicity in both cases).
Again, very similar results were obtained with other IMFs, extinction
recipes, and stellar population libraries. Under an
  unrealistic assumption of A(V)$=$0, the best-fitting ages were
  $t$$=$1.5~Gyr and $t$$=$2.6~Gyr for the low- and high-mass stacks,
  respectively. These solutions provide, based on the $\chi^2$ values,
  worse fits, and do not affect our main conclusions (cf.
  Section~\ref{sec:dis}).

Finally, we measured the stellar population ages fitting the whole
UV-to-NIR SED for each individual galaxy also using the Montecarlo
method, and calculating average properties for the low and high-mass
sub-samples. These were remarkably and reassuringly similar (within
the uncertainties) to the ones obtained with the other methods:
t$=$1.0$\pm$0.2~Gyr with A(V)$=$0.9$\pm$0.2~mag and
t$=$1.5$\pm$0.3~Gyr with A(V)$=$0.9$\pm$0.3~mag for the low-mass and
high-mass samples, respectively.

Our age estimations are completely consistent with those obtained by
\citet{Whitaker} using a stacked G141 grism spectrum around the
H$\beta$ absorption also for $UVJ$-selected MQGs, but at
1.4$<$z$<$2.2. They find ages between 0.9~Gyr and 1.6~Gyr for blue and
red massive galaxies, very similar to the ranges we find for the our
two sub-samples. Consistent ages are also found for MQGs at z$>$1
(selected in a variety of ways and counting with heterogeneous data)
by \citet{Onodera}, \citet{vds13}, \citet{bedregal13}, and
\citet{Marchesini}. In summary, the ages of the $UVJ$- and
sSFR-selected z$\sim$1 MQGs are confidently constrained to
be~$<$2~Gyr.

\subsection{IMF estimation} \label{sec:imf}

\begin{figure}[!htb]
    \includegraphics[width=0.5\textwidth]{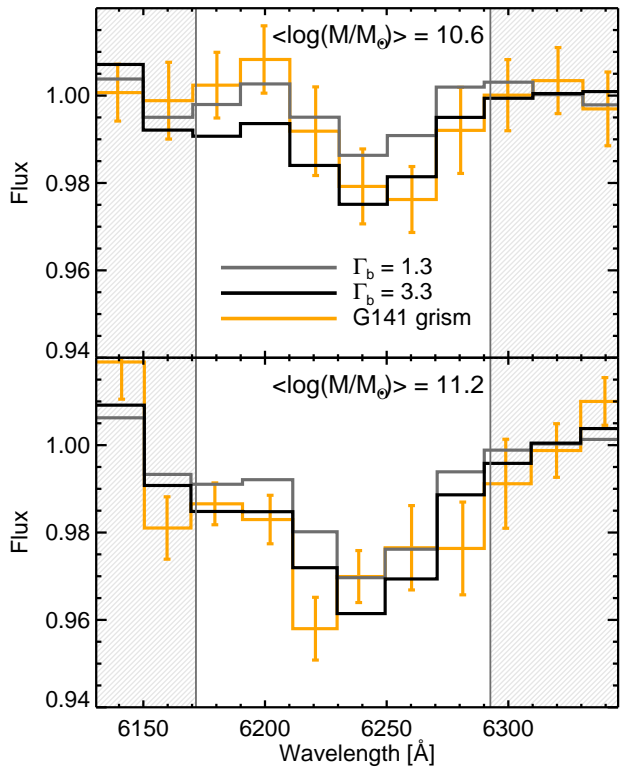} 
    
    \caption{The TiO$_2$ spectral region of the low- (top) and
      high-mass (bottom) stacks, as observed through the WFC3 G141
      grism (orange solid line). Data points are compared to
      models (smoothed to the same resolution) with a bottom-heavy
      (black histogram) and a standard Kroupa-like IMFs (gray). The
      observed spectra and models were normalized to the flux in the
      continuum bands (gray shaded regions). Ages were fixed to the
      results discussed in Section~\ref{sec:age}.}

    \label{fig:tio2}
\end{figure}

Once average ages were determined, we proceeded to the IMF analysis
based on the TiO$_2$ absorption. Given that this molecular band
dominates the spectrum of cool-dwarf stars between 600 and 640~nm, it
has been widely used to infer the IMF slope in unresolved stellar
systems \citep{ferreras,labarbera,Spiniello2013}.

We used the MILES SSP models \citep{miles}, where the IMF is
parametrized as a single power law, truncated (i.e., flatted out) for
stellar masses below M$<$0.6~M$_\odot$. This bimodal IMF is completely
described by a single parameter, $\Gamma_\mathrm{b}$
\citep[see][]{vazdekis96}. Under this parametrization, the Kroupa
(2001) IMF is recovered for $\Gamma_\mathrm{b}$$=$1.3.  The main
advantage of the bimodal IMF, compared to a regular single power law
(Salpeter-like) IMF, is the fact that, even when dealing with very
high $\Gamma_\mathrm{b}$ values, the $M/L$ ratio remains within the
observational limits suggested by dynamical studies \citep{ferreras}.
From the point of view of the stellar population properties, both
bimodal and uni-modal IMF parametrizations are indistinguishable.

MILES models cover a range from $-2.32$~dex to $+0.22$~dex in
metallicity, 0.06 Gyr to 17~Gyr in age, and
$\Gamma_\mathrm{b}$$=$0.3--3.3 in IMF slopes. Given the weak
dependence of the TiO$_2$ index with metallicity, we fixed it to solar
(as suggested by the SED fitting).

The classical definition for the TiO$_2$ spectral index expands along
$\sim$400~\AA, making it extremely sensitive to the adopted flux
calibration \citep[see Section~5 in][]{mn14}. To improve the signal,
we redefined the blue and red TiO$_2$ pseudo-continua, making them
contiguous to the central bandpass.  The adopted blue and red
pseudo-continua are 613.0--617.2~nm and 629.3--634.5~nm, respectively.
Figure~\ref{fig:tio2} presents, for both stacks, the
data and fits to the TiO$_2$ spectral region.

The analysis of the TiO$_2$ absorption was based on fits to the six
spectral elements (P$_\mathrm{obs}(\lambda)$) within the central band
of our TiO$_2$ index definition, after removing the continuum. The
models were degraded to the same spectral resolution 
(P$_\mathrm{SSP}(\lambda)$). The goodness of the fit was estimated with a 
$\chi^2$
function:

\begin{equation}
\chi^2(\Gamma_\mathrm{b}, \mathrm{age})=  \sum_\lambda
\frac{\phantom{^2}\left[ \mathrm{P}_\mathrm{obs}(\lambda) - 
\mathrm{P}_\mathrm{SSP}(\lambda)\right]^2}{\sigma^2_\mathrm{obs}(\lambda)}
\label{eq:chi}
\end{equation}

\noindent where $\sigma_\mathrm{obs}(\lambda)$ represents the
estimated error of the flux in each spectral bin. The $\chi^2$ maps in
the age-IMF slope plane for the low- and high-mass samples are shown
in Figure~\ref{fig:chi}.

\section{Discussion} \label{sec:dis}

\begin{figure}[!h]
 \begin{center}
  \includegraphics[width=0.5\textwidth]{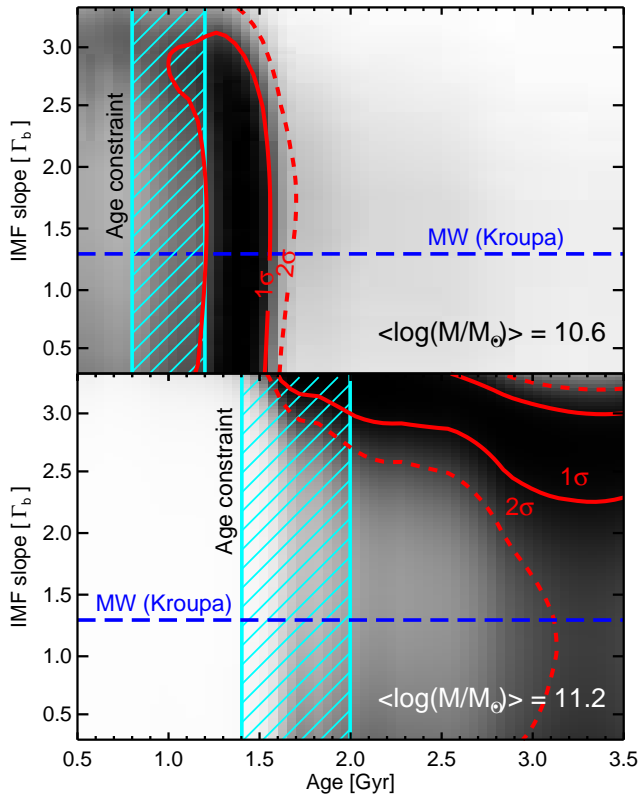}
 
  \caption{$\chi^2$ values in the IMF slope vs. age plane for the
    low-mass (top) and the high-mass (bottom) samples.  Darker tones
    indicate more probable SSP solutions. The solid and dashed red
    lines enclose the 1- and 2-$\sigma$ probability contours. Dashed 
    cyan regions mark the age range inferred from SED fitting.
    The combination of the TiO$_2$ index measurements and the stellar
    ages indicates that the IMF of massive quiescent galaxies at
    z$\sim$1 is bottom-heavy.  For the
    low-mass galaxies, degeneracies are larger and the IMF slope
    determination is significantly more uncertain.}

 \label{fig:chi}
 \end{center}
\end{figure}

Figure~\ref{fig:chi} shows our constraints on the stellar population
age and IMF slope for MQGs at z$\sim$1.  As expected, there is a clear
IMF/age degeneracy: similar TiO$_2$ values are obtained by either
an old population with a standard Kroupa-like IMF or with a
steeper IMF and younger ages. To further constrain the IMF, we use
the age determinations from the SED fitting.

For the high-mass sample, Figure~\ref{fig:chi} shows that our age
determination of 1.7$\pm$0.3~Gyr combined with the TiO$_2$ index
measurements strongly suggest that the IMF of
M$\gtrsim$10$^{11}$~M$_\sun$ MQGs at z$\sim$1 is bottom-heavy. The IMF
slope is $\Gamma_b$=3.2$\pm$0.2, very similar to that measured for
present-day early-type galaxies \citep{labarbera,Spiniello2013}.  For
the low-mass stack, considering a typical age of 1.0$\pm$0.2~Gyr, we
find that the IMF is flatter: $\Gamma_b=2.7^{+0.3}_{-0.4}$. The
uncertainty in this case is larger, mainly because the degeneracies
between age and IMF increase for younger ages and flatter IMFs. Using these IMF
values, and assuming a bimodal parametrization, the mass-limits of our stacks 
change to M$>$10$^{11.5}$M$_\sun$ and 10$^{10.7}$$<$M$<$10$^{11.5}$~M$_\sun$ 
for the high- and low-mass stacks, respectively. 
Although our age constraints are rather conservative (see
  Section~\ref{sec:age}), an offset of 0.5~Gyr in the lighter stack
  would leave the IMF slope unconstrained below $\Gamma_b\sim3$. Such
  a large error in a 1~Gyr old population is not expected, but the IMF
  determination of this lighter stack should be considered more
  tentative than that for the massive stack.  Furthermore, low-mass
galaxies tend to have more extended SFHs \citep{thomas05} and
therefore, their SED may be less well represented by a single SSP.
Note also that the departure from a SSP is expected to become larger
if galaxies are observed closer to their formation age.  This
  slightly extended star formation history in the lighter stack
  increases the scatter in the UV region, as shown in the upper left
  panel of Figures~\ref{data},\ref{data2}.  In addition, at lower
stellar masses the nature of galaxies becomes more heterogeneous,
increasing the likelihood of having systems following different
evolutionary tracks (e.g., disks and spheroids with different assembly
histories maybe affecting the IMF).

Two main caveats should be considered before further interpreting our
data: the effect of $\alpha$-element enhancement and metallicity.  Our
fits do not account for non-solar $\alpha$-elements abundances.
Massive galaxies exhibit an enhanced fraction of $\alpha$-elements
compared to the solar neighborhood, commonly interpreted as an imprint
of a fast formation process \citep{thomas05}. For a 1-2 Gyr old
population, an overabundance of $\sim$1~dex in [Ti/Fe] would be needed
to mimic the effect of a $\Gamma_\mathrm{b}$$=$3.2 IMF
\citep{thomas11b}. However, \citet{labarbera} found an excess of only
$\sim$0.2~dex in [Ti/Fe] for massive galaxies at $z\sim0$.
Therefore, unless the situation is totally different at high-z
\citep[but see][]{Choi}, our TiO$_2$ measurement is unlikely to be
explained with a standard IMF plus a non-solar [Ti/Fe] abundance. The
second caveat relates to the fact that we have used models with fixed
solar metallicity. The effect of the metallicity on the TiO$_2$ line
is very weak but not null. In this sense, we find steeper IMFs when
assuming larger metallicities. However, neither our SED fits, nor
z$\sim$0 massive galaxies \citep{labarbera} suggest a strong departure
from solar metallicity. On the contrary, an overestimation of
  the actual metallicity would weakly mimic the effect of a step IMF
  slope on the TiO$_2$ feature. However, sub-solar metallicities
  can be ruled out considering that galaxies as massive as
  those in our sample, show almost no metallicity evolution since
  $z\sim1$ \citep{Choi}, being metal-rich at $z\sim0$
  \citep{labarbera}. Thus, our results are robust against a poor 
  metallicity determination.

\begin{figure}[!htb]
  \begin{center}
    \includegraphics[width=0.48\textwidth,bb= 30 10 483 340,clip]{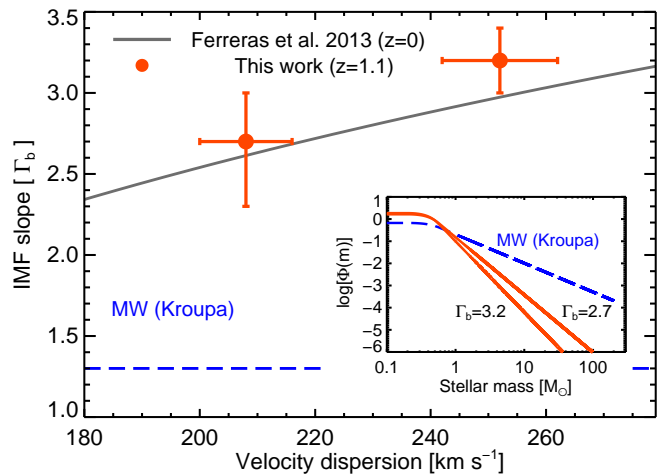} 
    \caption{IMF slope vs. velocity dispersion for MQGs at z$\sim$1,
      compared to the relation found for present-day ETGs
      \citep{ferreras} and a Kroupa (2001) IMF.  The inset explicitly
      shows the differences among all these IMFs.}
    \label{fig:rela}
  \end{center}
\end{figure}

In a more qualitative way, in Figure~\ref{fig:rela} we compare our
results with the IMF slope vs. velocity dispersion relation found in
the nearby Universe \citep{ferreras}. We have translated our stellar
mass scale to velocity dispersion using individual measurements for
our galaxies and statistical properties for samples at the same
redshift and selected in similar way. Based on measurements found in
the literature \citep[mainly in][]{vds13,belli} for galaxies
  with similar masses at similar redshifts, we obtain an average
velocity dispersion of 252$\pm$10~km\,s$^{-1}$ and
208$\pm$8~km\,s$^{-1}$ for our high- and low-mass stacks,
respectively. In addition, individual velocity dispersions have been
measured for two galaxies contributing to our low-mass stacked
spectrum \citep{Newman}. The mass of one of these galaxies is
M$=$10$^{10.6}$~M$_\sun$ and its velocity dispersion
$\sigma$$=$206~km\,s$^{-1}$, and for the other
M$=$10$^{10.9}$~M$_\sun$ and $\sigma$$=$239~km\,s$^{-1}$.  According
to Figure~\ref{fig:rela}, our z$\sim$1 IMF estimations are in good
agreement with the IMF slope in early-type galaxies of similar mass in
the present-day Universe. This suggests a direct evolutionary
  link between both populations and that the IMF, a key
characteristic of the stellar populations in galaxies, have remained
unchanged in the last $\sim$8 Gyr.

\acknowledgments 

We acknowledge support from the Spanish Government grants
AYA2010-21322-C03-02 and AYA2012-31277, and the ERC Advanced Grant
321323-NEOGAL. This work is based on SHARDS observations made with the
Gran Telescopio Canarias (GTC), and the Rainbow Cosmological Surveys
Database, operated by UCM partnered with UCO/Lick, UCSC. IMN thank
Carsten Weidner, Jes\'us Falc\'on-Barroso, and Mike Beasley for their
careful reading and comments on the manuscript.


\begin{thebibliography}{}
\expandafter\ifx\csname natexlab\endcsname\relax\def\natexlab#1{#1}\fi

\bibitem[{{Barro} {et~al.}(2011{\natexlab{a}}){Barro},
  {P{\'e}rez-Gonz{\'a}lez}, {Gallego}, {Ashby}, {Kajisawa}, {Miyazaki},
  {Villar}, {Yamada}, \& {Zamorano}}]{barro11a}
{Barro}, G., {P{\'e}rez-Gonz{\'a}lez}, P.~G., {Gallego}, J., {et~al.}
  2011{\natexlab{a}}, \apjs, 193, 13

\bibitem[{{Barro} {et~al.}(2011{\natexlab{b}}){Barro},
  {P{\'e}rez-Gonz{\'a}lez}, {Gallego}, {Ashby}, {Kajisawa}, {Miyazaki},
  {Villar}, {Yamada}, \& {Zamorano}}]{barro11b}
---. 2011{\natexlab{b}}, \apjs, 193, 30

\bibitem[{{Bastian} {et~al.}(2010){Bastian}, {Covey}, \& {Meyer}}]{bastian}
{Bastian}, N., {Covey}, K.~R., \& {Meyer}, M.~R. 2010, \araa, 48, 339

\bibitem[{{Bedregal} {et~al.}(2013){Bedregal}, {Scarlata}, {Henry}, {Atek},
  {Rafelski}, {Teplitz}, {Dominguez}, {Siana}, {Colbert}, {Malkan}, {Ross},
  {Martin}, {Dressler}, {Bridge}, {Hathi}, {Masters}, {McCarthy}, \&
  {Rutkowski}}]{bedregal13}
{Bedregal}, A.~G., {Scarlata}, C., {Henry}, A.~L., {et~al.} 2013, \apj, 778,
  126

\bibitem[{{Belli} {et~al.}(2014){Belli}, {Newman}, \& {Ellis}}]{belli}
{Belli}, S., {Newman}, A.~B., \& {Ellis}, R.~S. 2014, \apj, 783, 117

\bibitem[{{Bruzual} \& {Charlot}(2003)}]{bc03}
{Bruzual}, G., \& {Charlot}, S. 2003, \mnras, 344, 1000

\bibitem[{{Calzetti} {et~al.}(2000){Calzetti}, {Armus}, {Bohlin}, {Kinney},
  {Koornneef}, \& {Storchi-Bergmann}}]{Calzetti}
{Calzetti}, D., {Armus}, L., {Bohlin}, R.~C., {et~al.} 2000, \apj, 533, 682

\bibitem[{{Cappellari} {et~al.}(2012){Cappellari}, {McDermid}, {Alatalo},
  {Blitz}, {Bois}, {Bournaud}, {Bureau}, {Crocker}, {Davies}, {Davis}, {de
  Zeeuw}, {Duc}, {Emsellem}, {Khochfar}, {Krajnovi{\'c}}, {Kuntschner},
  {Lablanche}, {Morganti}, {Naab}, {Oosterloo}, {Sarzi}, {Scott}, {Serra},
  {Weijmans}, \& {Young}}]{cappellari}
{Cappellari}, M., {McDermid}, R.~M., {Alatalo}, K., {et~al.} 2012, \nat, 484,
  485

\bibitem[{{Cenarro} {et~al.}(2003){Cenarro}, {Gorgas}, {Vazdekis}, {Cardiel},
  \& {Peletier}}]{cenarro}
{Cenarro}, A.~J., {Gorgas}, J., {Vazdekis}, A., {Cardiel}, N., \& {Peletier},
  R.~F. 2003, \mnras, 339, L12

\bibitem[{{Chabrier}(2003)}]{chabrier}
{Chabrier}, G. 2003, \pasp, 115, 763

\bibitem[{{Charlot} \& {Fall}(2000)}]{Charlot00}
{Charlot}, S., \& {Fall}, S.~M. 2000, \apj, 539, 718

\bibitem[{{Choi} {et~al.}(2014){Choi}, {Conroy}, {Moustakas}, {Graves},
  {Holden}, {Brodwin}, {Brown}, \& {van Dokkum}}]{Choi}
{Choi}, J., {Conroy}, C., {Moustakas}, J., {et~al.} 2014, ArXiv e-prints,
  arXiv:1403.4932

\bibitem[{{Conroy} {et~al.}(2013){Conroy}, {Dutton}, {Graves}, {Mendel}, \&
  {van Dokkum}}]{Conroy13}
{Conroy}, C., {Dutton}, A.~A., {Graves}, G.~J., {Mendel}, J.~T., \& {van
  Dokkum}, P.~G. 2013, \apjl, 776, L26

\bibitem[{{Dav{\'e}}(2008)}]{dave}
{Dav{\'e}}, R. 2008, \mnras, 385, 147

\bibitem[{{Ferreras} {et~al.}(2013{\natexlab{a}}){Ferreras}, {La Barbera}, {de
  la Rosa}, {Vazdekis}, {de Carvalho}, {Falc{\'o}n-Barroso}, \&
  {Ricciardelli}}]{ferreras}
{Ferreras}, I., {La Barbera}, F., {de la Rosa}, I.~G., {et~al.}
  2013{\natexlab{a}}, \mnras, 429, L15

\bibitem[{{Ferreras} {et~al.}(2013{\natexlab{b}}){Ferreras}, {Trujillo},
  {M{\'a}rmol-Queralt{\'o}}, {P{\'e}rez-Gonz{\'a}lez}, {Cava}, {Barro},
  {Cenarro}, {Hern{\'a}n-Caballero}, \& {Cardiel}}]{ferreras14}
{Ferreras}, I., {Trujillo}, I., {M{\'a}rmol-Queralt{\'o}}, E., {et~al.}
  2013{\natexlab{b}}, ArXiv e-prints, arXiv:1312.5317

\bibitem[{{Grogin} {et~al.}(2011){Grogin}, {Kocevski}, {Faber}, {Ferguson},
  {Koekemoer}, {Riess}, {Acquaviva}, {Alexander}, {Almaini}, {Ashby}, {Barden},
  {Bell}, {Bournaud}, {Brown}, {Caputi}, {Casertano}, {Cassata}, {Castellano},
  {Challis}, {Chary}, {Cheung}, {Cirasuolo}, {Conselice}, {Roshan Cooray},
  {Croton}, {Daddi}, {Dahlen}, {Dav{\'e}}, {de Mello}, {Dekel}, {Dickinson},
  {Dolch}, {Donley}, {Dunlop}, {Dutton}, {Elbaz}, {Fazio}, {Filippenko},
  {Finkelstein}, {Fontana}, {Gardner}, {Garnavich}, {Gawiser}, {Giavalisco},
  {Grazian}, {Guo}, {Hathi}, {H{\"a}ussler}, {Hopkins}, {Huang}, {Huang},
  {Jha}, {Kartaltepe}, {Kirshner}, {Koo}, {Lai}, {Lee}, {Li}, {Lotz}, {Lucas},
  {Madau}, {McCarthy}, {McGrath}, {McIntosh}, {McLure}, {Mobasher},
  {Moustakas}, {Mozena}, {Nandra}, {Newman}, {Niemi}, {Noeske}, {Papovich},
  {Pentericci}, {Pope}, {Primack}, {Rajan}, {Ravindranath}, {Reddy}, {Renzini},
  {Rix}, {Robaina}, {Rodney}, {Rosario}, {Rosati}, {Salimbeni}, {Scarlata},
  {Siana}, {Simard}, {Smidt}, {Somerville}, {Spinrad}, {Straughn}, {Strolger},
  {Telford}, {Teplitz}, {Trump}, {van der Wel}, {Villforth}, {Wechsler},
  {Weiner}, {Wiklind}, {Wild}, {Wilson}, {Wuyts}, {Yan}, \& {Yun}}]{Grogin}
{Grogin}, N.~A., {Kocevski}, D.~D., {Faber}, S.~M., {et~al.} 2011, \apjs, 197,
  35

\bibitem[{{Hern{\'a}n-Caballero} {et~al.}(2013){Hern{\'a}n-Caballero},
  {Alonso-Herrero}, {P{\'e}rez-Gonz{\'a}lez}, {Cardiel}, {Cava}, {Ferreras},
  {Barro}, {Tresse}, {Daddi}, {Cenarro}, {Conselice}, {Guzm{\'a}n}, \&
  {Gallego}}]{hernan13}
{Hern{\'a}n-Caballero}, A., {Alonso-Herrero}, A., {P{\'e}rez-Gonz{\'a}lez},
  P.~G., {et~al.} 2013, \mnras, 434, 2136

\bibitem[{{Hoversten} \& {Glazebrook}(2008)}]{Hoversten}
{Hoversten}, E.~A., \& {Glazebrook}, K. 2008, \apj, 675, 163

\bibitem[{{Koekemoer} {et~al.}(2011){Koekemoer}, {Faber}, {Ferguson}, {Grogin},
  {Kocevski}, {Koo}, {Lai}, {Lotz}, {Lucas}, {McGrath}, {Ogaz}, {Rajan},
  {Riess}, {Rodney}, {Strolger}, {Casertano}, {Castellano}, {Dahlen},
  {Dickinson}, {Dolch}, {Fontana}, {Giavalisco}, {Grazian}, {Guo}, {Hathi},
  {Huang}, {van der Wel}, {Yan}, {Acquaviva}, {Alexander}, {Almaini}, {Ashby},
  {Barden}, {Bell}, {Bournaud}, {Brown}, {Caputi}, {Cassata}, {Challis},
  {Chary}, {Cheung}, {Cirasuolo}, {Conselice}, {Roshan Cooray}, {Croton},
  {Daddi}, {Dav{\'e}}, {de Mello}, {de Ravel}, {Dekel}, {Donley}, {Dunlop},
  {Dutton}, {Elbaz}, {Fazio}, {Filippenko}, {Finkelstein}, {Frazer}, {Gardner},
  {Garnavich}, {Gawiser}, {Gruetzbauch}, {Hartley}, {H{\"a}ussler},
  {Herrington}, {Hopkins}, {Huang}, {Jha}, {Johnson}, {Kartaltepe},
  {Khostovan}, {Kirshner}, {Lani}, {Lee}, {Li}, {Madau}, {McCarthy},
  {McIntosh}, {McLure}, {McPartland}, {Mobasher}, {Moreira}, {Mortlock},
  {Moustakas}, {Mozena}, {Nandra}, {Newman}, {Nielsen}, {Niemi}, {Noeske},
  {Papovich}, {Pentericci}, {Pope}, {Primack}, {Ravindranath}, {Reddy},
  {Renzini}, {Rix}, {Robaina}, {Rosario}, {Rosati}, {Salimbeni}, {Scarlata},
  {Siana}, {Simard}, {Smidt}, {Snyder}, {Somerville}, {Spinrad}, {Straughn},
  {Telford}, {Teplitz}, {Trump}, {Vargas}, {Villforth}, {Wagner}, {Wandro},
  {Wechsler}, {Weiner}, {Wiklind}, {Wild}, {Wilson}, {Wuyts}, \&
  {Yun}}]{Koekemoer}
{Koekemoer}, A.~M., {Faber}, S.~M., {Ferguson}, H.~C., {et~al.} 2011, \apjs,
  197, 36

\bibitem[{{Kroupa}(2001)}]{mw}
{Kroupa}, P. 2001, \mnras, 322, 231

\bibitem[{{Kroupa}(2002)}]{kroupa}
---. 2002, Science, 295, 82

\bibitem[{{Kroupa} {et~al.}(2013){Kroupa}, {Weidner}, {Pflamm-Altenburg},
  {Thies}, {Dabringhausen}, {Marks}, \& {Maschberger}}]{kroupa13}
{Kroupa}, P., {Weidner}, C., {Pflamm-Altenburg}, J., {et~al.} 2013, 115

\bibitem[{{La Barbera} {et~al.}(2013){La Barbera}, {Ferreras}, {Vazdekis}, {de
  la Rosa}, {de Carvalho}, {Trevisan}, {Falc{\'o}n-Barroso}, \&
  {Ricciardelli}}]{labarbera}
{La Barbera}, F., {Ferreras}, I., {Vazdekis}, A., {et~al.} 2013, \mnras, 433,
  3017

\bibitem[{{Le Borgne} {et~al.}(2003){Le Borgne}, {Bruzual}, {Pell{\'o}},
  {Lan{\c c}on}, {Rocca-Volmerange}, {Sanahuja}, {Schaerer}, {Soubiran}, \&
  {V{\'{\i}}lchez-G{\'o}mez}}]{stelib}
{Le Borgne}, J.-F., {Bruzual}, G., {Pell{\'o}}, R., {et~al.} 2003, \aap, 402,
  433

\bibitem[{{Marchesini} {et~al.}(2014){Marchesini}, {Muzzin}, {Stefanon},
  {Franx}, {Brammer}, {Marsan}, {Vulcani}, {Fynbo}, {Milvang-Jensen}, {Dunlop},
  \& {Buitrago}}]{Marchesini}
{Marchesini}, D., {Muzzin}, A., {Stefanon}, M., {et~al.} 2014, ArXiv e-prints,
  arXiv:1402.0003

\bibitem[{{Mart{\'{\i}}n-Navarro} {et~al.}(2014){Mart{\'{\i}}n-Navarro}, {La
  Barbera}, {Vazdekis}, {Falc{\'o}n-Barroso}, \& {Ferreras}}]{mn14}
{Mart{\'{\i}}n-Navarro}, I., {La Barbera}, F., {Vazdekis}, A.,
  {Falc{\'o}n-Barroso}, J., \& {Ferreras}, I. 2014, ArXiv e-prints,
  arXiv:1404.6533

\bibitem[{{Meurer} {et~al.}(1999){Meurer}, {Heckman}, \& {Calzetti}}]{Meurer}
{Meurer}, G.~R., {Heckman}, T.~M., \& {Calzetti}, D. 1999, \apj, 521, 64

\bibitem[{{Meurer} {et~al.}(2009){Meurer}, {Wong}, {Kim}, {Hanish}, {Heckman},
  {Werk}, {Bland-Hawthorn}, {Dopita}, {Zwaan}, {Koribalski}, {Seibert},
  {Thilker}, {Ferguson}, {Webster}, {Putman}, {Knezek}, {Doyle}, {Drinkwater},
  {Hoopes}, {Kilborn}, {Meyer}, {Ryan-Weber}, {Smith}, \&
  {Staveley-Smith}}]{Meurer09}
{Meurer}, G.~R., {Wong}, O.~I., {Kim}, J.~H., {et~al.} 2009, \apj, 695, 765

\bibitem[{{Mould}(1976)}]{tio2}
{Mould}, J.~R. 1976, \aap, 48, 443

\bibitem[{{Newman} {et~al.}(2010){Newman}, {Ellis}, {Treu}, \&
  {Bundy}}]{Newman}
{Newman}, A.~B., {Ellis}, R.~S., {Treu}, T., \& {Bundy}, K. 2010, \apjl, 717,
  L103

\bibitem[{{Onodera} {et~al.}(2012){Onodera}, {Renzini}, {Carollo},
  {Cappellari}, {Mancini}, {Strazzullo}, {Daddi}, {Arimoto}, {Gobat}, {Yamada},
  {McCracken}, {Ilbert}, {Capak}, {Cimatti}, {Giavalisco}, {Koekemoer}, {Kong},
  {Lilly}, {Motohara}, {Ohta}, {Sanders}, {Scoville}, {Tamura}, \&
  {Taniguchi}}]{Onodera}
{Onodera}, M., {Renzini}, A., {Carollo}, M., {et~al.} 2012, \apj, 755, 26

\bibitem[{{P{\'e}rez-Gonz{\'a}lez} {et~al.}(2008){P{\'e}rez-Gonz{\'a}lez},
  {Rieke}, {Villar}, {Barro}, {Blaylock}, {Egami}, {Gallego}, {Gil de Paz},
  {Pascual}, {Zamorano}, \& {Donley}}]{pablo08}
{P{\'e}rez-Gonz{\'a}lez}, P.~G., {Rieke}, G.~H., {Villar}, V., {et~al.} 2008,
  \apj, 675, 234

\bibitem[{{P{\'e}rez-Gonz{\'a}lez} {et~al.}(2013){P{\'e}rez-Gonz{\'a}lez},
  {Cava}, {Barro}, {Villar}, {Cardiel}, {Ferreras},
  {Rodr{\'{\i}}guez-Espinosa}, {Alonso-Herrero}, {Balcells}, {Cenarro}, {Cepa},
  {Charlot}, {Cimatti}, {Conselice}, {Daddi}, {Donley}, {Elbaz}, {Espino},
  {Gallego}, {Gobat}, {Gonz{\'a}lez-Mart{\'{\i}}n}, {Guzm{\'a}n},
  {Hern{\'a}n-Caballero}, {Mu{\~n}oz-Tu{\~n}{\'o}n}, {Renzini},
  {Rodr{\'{\i}}guez-Zaur{\'{\i}}n}, {Tresse}, {Trujillo}, \&
  {Zamorano}}]{shards}
{P{\'e}rez-Gonz{\'a}lez}, P.~G., {Cava}, A., {Barro}, G., {et~al.} 2013, \apj,
  762, 46

\bibitem[{{Renzini}(2006)}]{renzini}
{Renzini}, A. 2006, \araa, 44, 141

\bibitem[{{Salpeter}(1955)}]{salpeter}
{Salpeter}, E.~E. 1955, \apj, 121, 161

\bibitem[{{Shetty} \& {Cappellari}(2014)}]{Shetty}
{Shetty}, S., \& {Cappellari}, M. 2014, \apjl, 786, L10

\bibitem[{{Spiniello} {et~al.}(2014){Spiniello}, {Trager}, {Koopmans}, \&
  {Conroy}}]{Spiniello2013}
{Spiniello}, C., {Trager}, S., {Koopmans}, L.~V.~E., \& {Conroy}, C. 2014,
  \mnras, 438, 1483

\bibitem[{{Thomas} {et~al.}(2005){Thomas}, {Maraston}, {Bender}, \& {Mendes de
  Oliveira}}]{thomas05}
{Thomas}, D., {Maraston}, C., {Bender}, R., \& {Mendes de Oliveira}, C. 2005,
  \apj, 621, 673

\bibitem[{{Thomas} {et~al.}(2011){Thomas}, {Maraston}, \&
  {Johansson}}]{thomas11b}
{Thomas}, D., {Maraston}, C., \& {Johansson}, J. 2011, \mnras, 412, 2183

\bibitem[{{Treu} {et~al.}(2010){Treu}, {Auger}, {Koopmans}, {Gavazzi},
  {Marshall}, \& {Bolton}}]{treu}
{Treu}, T., {Auger}, M.~W., {Koopmans}, L.~V.~E., {et~al.} 2010, \apj, 709,
  1195

\bibitem[{{van de Sande} {et~al.}(2013){van de Sande}, {Kriek}, {Franx}, {van
  Dokkum}, {Bezanson}, {Bouwens}, {Quadri}, {Rix}, \& {Skelton}}]{vds13}
{van de Sande}, J., {Kriek}, M., {Franx}, M., {et~al.} 2013, \apj, 771, 85

\bibitem[{{van Dokkum}(2008)}]{vd08}
{van Dokkum}, P.~G. 2008, \apj, 674, 29

\bibitem[{{van Dokkum} \& {Conroy}(2010)}]{vandokkum}
{van Dokkum}, P.~G., \& {Conroy}, C. 2010, \nat, 468, 940

\bibitem[{{Vazdekis} {et~al.}(1996){Vazdekis}, {Casuso}, {Peletier}, \&
  {Beckman}}]{vazdekis96}
{Vazdekis}, A., {Casuso}, E., {Peletier}, R.~F., \& {Beckman}, J.~E. 1996,
  \apjs, 106, 307

\bibitem[{{Vazdekis} {et~al.}(2010){Vazdekis}, {S{\'a}nchez-Bl{\'a}zquez},
  {Falc{\'o}n-Barroso}, {Cenarro}, {Beasley}, {Cardiel}, {Gorgas}, \&
  {Peletier}}]{miles}
{Vazdekis}, A., {S{\'a}nchez-Bl{\'a}zquez}, P., {Falc{\'o}n-Barroso}, J.,
  {et~al.} 2010, \mnras, 404, 1639

\bibitem[{{Whitaker} {et~al.}(2011){Whitaker}, {Labb{\'e}}, {van Dokkum},
  {Brammer}, {Kriek}, {Marchesini}, {Quadri}, {Franx}, {Muzzin}, {Williams},
  {Bezanson}, {Illingworth}, {Lee}, {Lundgren}, {Nelson}, {Rudnick}, {Tal}, \&
  {Wake}}]{Whitaker2011}
{Whitaker}, K.~E., {Labb{\'e}}, I., {van Dokkum}, P.~G., {et~al.} 2011, \apj,
  735, 86

\bibitem[{{Whitaker} {et~al.}(2013){Whitaker}, {van Dokkum}, {Brammer},
  {Momcheva}, {Skelton}, {Franx}, {Kriek}, {Labb{\'e}}, {Fumagalli},
  {Lundgren}, {Nelson}, {Patel}, \& {Rix}}]{Whitaker}
{Whitaker}, K.~E., {van Dokkum}, P.~G., {Brammer}, G., {et~al.} 2013, \apjl,
  770, L39

\end{thebibliography}

\end{document}